\documentclass [global,twocolumn]{svjour}
\usepackage[dvips]{graphics}
\usepackage {longtable}
\usepackage{t1enc}
\usepackage[english]{babel}
\usepackage{epsfig}
\usepackage{amssymb}

\title{Optimizing the production of metastable calcium atoms in a
magneto-optical trap}
\author{ Jan Gr\"{u}nert \and Andreas Hemmerich}
\institute{Institut für Laserphysik, Universit\"{a}t Hamburg,
Jungiusstrasse 9, D-20355 Hamburg, Germany
\\ \email{gruenert@physnet.uni-hamburg.de}}

\begin{document}

\maketitle

\abstract {We investigate the production of long lived metastable
(4$^3$P$_2$) calcium atoms in a magneto-optical trap operating on
the 4$^1$S$_0  \to $ 4$^1$P$_1$ transition at 423~nm. For excited
4$^1$P$_1$~atoms a weak decay channel into the triplet states
4$^3$P$_2$ and 4$^3$P$_1$ exists via the 3$^1$D$_2$~state. The
undesired 4$^3$P$_1$~atoms decay back to the ground state within
0.4 ms and can be fully recaptured if the illuminated trap volume
is sufficiently large. We obtain a flux of above $10^{10}$ atoms/s
into the 4$^3$P$_2$~state. We find that our MOT life time of 23~ms
is mainly limited by this loss channel and thus the 4$^{3}$P$_2$
production is not hampered by inelasic collisions. If we close the
loss channel by repumping the 3$^1$D$_2 $~atoms with a 671~nm
laser back into the MOT cycling transition, a non-exponential
72~ms trap decay is observed indicating the presence of inelastic
two-body collisions between 4$^1$S$_{0}$ and 4$^1$P$_1 $atoms.

}

\section{Introduction}

\begin{figure}
  \centering
  \scalebox{0.7}{\includegraphics{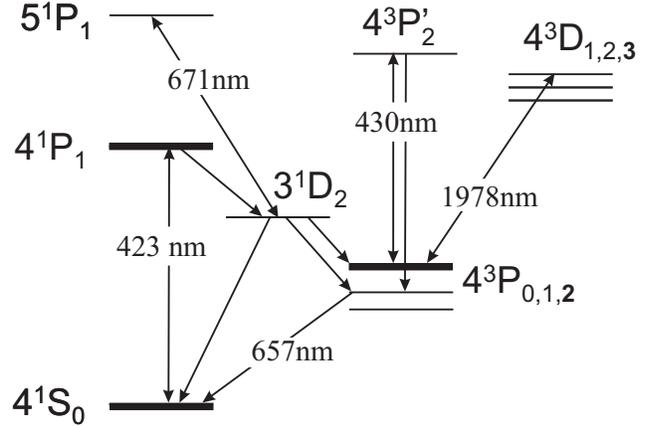}}
  \caption{Relevant energy levels of $^{40}$Ca including the strong cooling and
trapping line at 423~nm, the repumping laser at 671~nm, the
cooling and probing transitions for metastable atoms at 1978~nm
and 430~nm respectively. }\label{levels}
\end{figure}

In recent years the development of magneto-optical traps (MOT)
proceeded upon the detailed investigation of laser cooling and
trapping in the alkali group, that culminated in the observation
of Bose-Einstein condensation (BEC) \cite{anderson,davis}.
Although selecting alkali atoms for this venture proved successful
due to the simplicity of their energy level schemes, it became
obvious, that in order to shorten the cooling times towards
condensation of typically > 10 s one would have to abandon the
concept of evaporative cooling in magnetic traps and proceed to
all-optical solutions instead. Earth-alkali atoms offer narrow
optical transition lines because of their two valence electrons
and consequent energy level complexity of spin zero singlet states
and spin one triplet states. These intercombination lines open up
new possibilities for improved cooling schemes with the promise to
reach particularly low temperatures close to or even below the
critical temperature for BEC \cite{katori,PTB,NIST}. Novel,
possibly very efficient paths towards quantum degenracy as, for
example, matter wave amplification by optical pumping (MAO)
\cite{spreeuw} may become feasible with earth alkaline atoms
\cite{gruenert}. Our approach to a realization of the MAO scheme
in calcium involves a dense and cold MOT for long-lived (118~min.
\cite{derev}) metastable 4$^3$P$_2$~atoms \cite{PRL}. This MOT
operates on the closed cycle 4$^3$P$_2  \to $ 4$^3$D$_3$
transition at 1978~nm (cf. fig. \ref{levels}).
Because of its narrow line width of 60~kHz this infrared
transition provides a Doppler-temperature of only 3~microkelvin.
In addition, its Zeeman substructure implies the presence of
polarization gradient cooling with the promise of temperatures
approaching the recoil limit of 122~nK. Such cold calcium samples
would yield a wealth of new physics. For example, the simple
structure of the zero spin ground state allows for precise
comparisons between theory and experiments in ultracold collision
studies. Ultracold calcium samples should also boost the
developement of novel optical time and length standards based on
the 408~Hz intercombination line at 657~nm \cite{kisters}.

\section{Production scheme of metastable calcium samples}

In order to load a 4$^3$P$_2$-MOT, precooled 4$^3$P$_2$~atoms have
to be produced at a high rate. In this paper we discuss the
production of a continuous flux of more than 10$^{10}$ cold
4$^3$P$_2$~atoms per second from a MOT operating on the 4$^1$S$_0
\to $ 4$^1$P$_1$ transition at 423~nm (natural linewidth
$\Gamma_{nat} /2\pi  = 34.6$~MHz). We utilize the fact that there
is a weak decay channel for excited 4$^1$P$_1$~atoms into the
3$^1$D$_2 $~state occuring at a rate of 2180~s$^{-1}$
\cite{strumia}. About 78 \% of the 3$^1$D$_2$~atoms return to the
ground state in approximately 3~ms either directly in a quadrupole
transition or via the 4$^3$P$_1$~state which decays via the 657~nm
intercombination line. These atoms are not subject to trapping
forces during this time and will move out of the trap at an
average speed of 1~m/s. Thus the illuminated capture volume of the
MOT has to be sufficiently large (Ø = 1 cm) to completely
recapture them once they returned to the ground state. The
remaining 22 \% are transfered to the desired metastable
4$^3$P$_2$~state. These atoms exhibit the Doppler-limited MOT
temperature of typically 1-2~mK, which appears ideal as a starting
point for further cooling.

\section{Experimental apparatus}

A detailed description of the 4$^1$S$_0 $-MOT setup is given in
\cite{gruenert}. This MOT is loaded from a thermal Ca oven at
650$^\circ$C through a conventional decreasing field Zeeman slower
operated with a slowing laser tuned 270~MHz below resonance. The
appropriate laser light at 423~nm is produced by a
frequency-doubled Ti:Sapphire laser pumped with 10~W pumping power
at 532~nm from a commercial solid state laser system. This results
in about 300~mW of blue radiation of which 2/3 can be used for the
experiment, having to be shared upon the Zeeman cooler, a
two-dimensional transverse cooling stage at its exit and the MOT
itself. The MOT-beams are about 10~mm in diameter yielding a
saturation parameter of 1/2. For optical pumping experiments
(repumping 3$^1$D$_2$-atoms) described below, we employ an
external cavity diode laser system at 671~nm. The 10 mW diode
provides 3.6 mW output at the 3$^1$D$_2 \to $ 5$^1$P$_1$
transition wavelength of 671.769~nm (air). A fraction of 1.3~mW is
available in the trap after beam shaping. For direct detection of
metastable 4$^3$P$_2$~atoms we use a frequency doubled external
cavity diode laser at 860~nm with 30~mW usable output power. After
resonant frequency doubling in KNbO$_3$ and beam shaping we have
5~mW available resonant with the 4$^3$P$_2$ (4s4p)$ \to $
4$^3$P$_2$ (4p$^2$)~transition at 430.252~nm (air). As a frequency
reference for this laser we have combined a DC discharge with a
calcium heatpipe. Inside a 15~cm glass tube calcium is heated to
600$^\circ$C in a small steel cup which represents the cathode of
a DC discharge operated with 1 kV in 2 torr of neon. Using
Doppler-free polarization spectroscopy we obtain dispersive
signals with signal to noise above 50, well suited for frequency
stabilization of the laser. For analyzing our system we can
observe independently the fluorescence at 657~nm with a photo
multiplier tube and at 423~nm with a calibrated photo diode.

\section{Flux of metastable atoms}

We can determine the flux of metastable atoms indirectly by
measuring the number of 4$^1$P$_1$~atoms in our MOT, which can be
achieved by observing the 423~nm fluorescence. In the steady state
we find a 4$^1$P$_1$-population N(4$^1$P$_1)=4\cdot10^{7}$ which
translates into a flux of
\begin{eqnarray*}
\mbox{N}(4^1\mbox{P}_1)~\cdot~0.22 ~\cdot~2180~\mbox{s}^{-1} = 2
\cdot10^{10}~\mbox{s}^{-1}
\end{eqnarray*}
\noindent into the 4$^3$P$_2$~state. We can check whether
N(4$^1$P$_1$) is limited by imperfect recycling of atoms decaying
from 3$^1$D$_2$ to the ground state by means of a life time
measurement of the MOT.
\begin{figure}
  \centering
  \scalebox{0.5}{\includegraphics{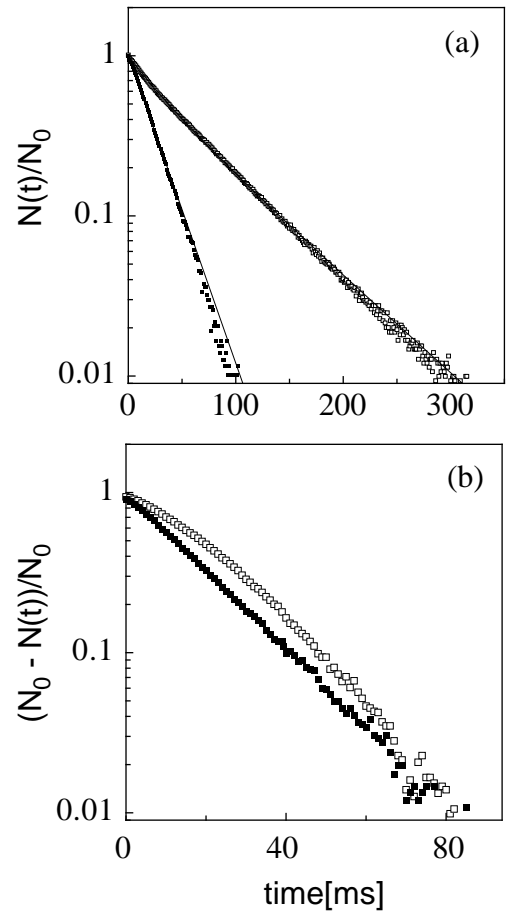}}
  \caption{Decay (a) and loading (b) of the trap with ($\square$) and
  without ($\blacksquare$) application of a
repumping laser acting on the $^1$D$_2$ population.}\label{decay}
\end{figure}
Such a measurement is done by shutting off
the Zeeman cooler, thus terminating further loading and observing
the decaying 423~nm fluorescence. The result in fig. \ref{decay}
shows a clean exponential decay with a time constant of 22.6 ms.
Using the rate equation model discussed in \cite{gruenert} (cf.
eq.(3)) with the saturation parameter of 0.5 in our MOT we obtain
100 \% recapture efficiency.

\begin{figure}
  \centering
  \scalebox{0.45}{\includegraphics{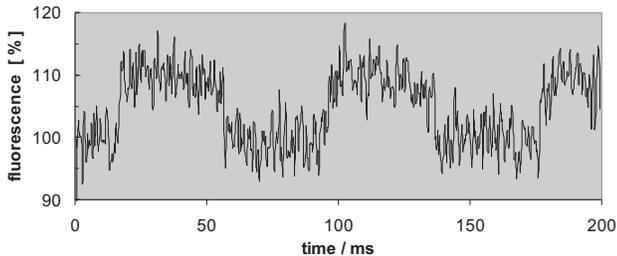}}
  \caption{When the 430~nm probing laser is pulsed on while the MOT is operating
  the red fluorescence at 657~nm is increased by 12 \% because metastable $^3$P$_2$ atoms
  are transfered to $^3$P$_1$ and decay to the ground state via the intercombination line.}
  \label{pulsing}
\end{figure}

In order to observe the flux of atoms into the 4$^3$P$_2$-state in
a direct fashion we have employed an additional laser beam at
430~nm which optically pumps all 4$^3$P$_2$~atoms into the
4$^3$P$_1$~state with a rate above 10$^{8}$~s$^{-1}$. These
4$^3$P$_1$~atoms add to those produced by decay from 3$^1$D$_2$
thus increasing the 657~nm fluorescence emerging from the MOT when
they decay to the ground state. As illustrated in fig.
\ref{pulsing}
we observe an increase of the red fluorescence level by 1/8 when
the optical pumping laser is active. Under the realistic
assumption that all 4$^3$P$_2$~atoms are optically pumped to
4$^3$P$_1$ instantaneously with respect to the time scale of their
production our observation suggests, that the 3$^1$D$_2 \to $
4$^3$P$_2$~decay rate is 1/8 times that of the 3$^1$D$_2  \to $
4$^3$P$_1$~decay rate which is to be compared with the theoretical
branching ratio of 1/3. This discrepancy is resolved by noting
that the 3$^1$D$_{2 } \to $ 4$^3$P$_2$~decay takes 10~ms on
average and thus these atoms have traveled about 1~cm, three times
more than the 4$^3$P$_1$~atoms. As a consequence the
4$^3$P$_1$~atoms remain entirely in the region mapped onto the
photo multiplier, while this is only the case for about 40 \% of
the 4$^3$P$_2$~atoms.

\section{Possible limitations by cold collisions}

The purely exponential trap decay suggests that at the present
capture rate of the MOT collisional loss is not yet a limiting
loss mechanism comparable to the desired loss due to the
production of metastables. As a consequence, in our case of
perfect recycling the production rate of 4$^3$P$_2$~atoms equals
the capture rate of the 4$^1$S$_0 $-MOT, i.e all atoms captured
from the Zeeman cooler are eventually transfered into the desired
metastable state. It appears natural to ask whether a further
increase of the MOT capture rate by improvement of the Zeeman
cooler will yield a further increase of 4$^3$P$_2$-production, or
rather take us into the regime where inelastic collisions
contribute a comparable loss. In order to investigate this
question we have decreased the loss channel which results from the
production of metastables by repumping the 3$^1$D$_2 $~atoms with
a 671~nm laser back into the MOT cycling transition\cite{oates}.
In this case, we observe a more than three-fold increase of the
trap decay time to 72~ms (cf. fig. \ref{decay}a) in combination
with a twofold increased 423~nm fluorescence level. Moreover, the
decay acquires a clearly non-exponential character yielding a
slight curvature in the logarithmic plot of the trap decay in fig.
\ref{decay}(a). We have also observed the loading process of the
MOT by suddenly turning on the Zeeman cooler and recording the
423~nm fluorescence. It appears, that in contrast to the
observations regarding the trap decay there is only a very small
increase of the loading time constant if we activate the 671~nm
repumper. These observations strongly indicate the presence of an
additional density dependent loss channel occuring for the
repumped MOT. In absence of such an additional loss mechanism our
trap should be limited by the linear loss resulting from
collisions with hot background atoms, which for our imperfect
vacuum conditions would still suggest a more than tenfold increase
of the lifetime and the fluorescence level. Moreover, for linear
loss the loading and decay should be purely exponential with the
same time constant. We assume that inelastic two-body collisions
between cold 4$^1$S$_0 $ and 4$^1$P$_1$ atoms are responsible for
the addtional trap loss. This hypothesis is supported by
observations in Strontium, where a comparably large crossection
for such collisions has been observed and theoretically explained
with the exsistence of a metastable $^1\Pi _g$ molecular state for
the strontium dimer, which enables a close approach of the
4$^1$S$_0 $ and 4$^1$P$_1$ collision partners without spontaneous
emission \cite{SrSr}.

The rate equation for the trapping dynamics of the MOT including
two-body collisions reads
\begin{equation}
\mbox{\.{N}} =  R-\Gamma \mbox{N} - \beta \int n^2(r)\,d^3r
\end{equation}
where $R$ is the MOT capture rate, $\Gamma$ accounts for the
linear loss due to collisions with hot background atoms, $\beta$
is the coefficient for inelastic collisions and $n(r) = n
\,e^{{-\left(r/a\right)}^2}$ is the Gaussian distribution of the
atomic density. The solution of this differential equation for the
condition of terminated loading (trap decay) is

\begin{equation}\label{decay1}
  \frac{\mbox{N}}{\mbox{N}_0}=\frac{e^{-\Gamma t}}
  {1+\frac{\xi}{1-\xi}\left(1-e^{-\Gamma t}\right)}
\end{equation}
\noindent where $N_0=n_0\left(\sqrt{\pi}a\right)^3$ denotes the
steady state population,
\begin{equation}\label{n0}
n_0=\sqrt{\frac{2\Gamma^2}{\beta^2}+\frac{8R}{\beta}\left(\sqrt{2\pi}a\right)^3}
-\frac{\sqrt{8}\Gamma}{2\beta}
\end{equation}
\noindent is the corresponding steady state peak density in the
trap and
\begin{equation}\label{xi}
  \xi=\frac{\beta n_0}{\sqrt{8}\Gamma+\beta n_0}=\frac{\sqrt{1+\frac{4 R \beta}
  {\Gamma^2\left(\sqrt{2\pi}a\right)^3}}-1}{\sqrt{1+\frac{4 R \beta}
  {\Gamma^2\left(\sqrt{2\pi}a\right)^3}}+1}
\end{equation}
\noindent is a parameter in the interval $\left[0,1\right]$ which
denotes the fraction of quadratic loss relative to the total trap
loss.

In the case of trap loading the trapped population increases
according to

\begin{equation}\label{loading}
  1-\frac{\mbox{N}}{\mbox{N}_0}=\frac{e^{-\gamma t}}
  {\frac{1}{1+\xi}+\frac{\xi}{1+\xi} e^{-\gamma t}} \qquad , \quad
   \gamma=\frac{1+\xi}{1-\xi}\Gamma
\end{equation}

While the trap decay involves the linear loss rate $\Gamma $
loading occurs with an increased rate $\gamma $. The reason for
this behavior is that the loading is abruptly terminated once the
quadratic regime is reached leading to shorter loading times as
compared to what is expected due to linear losses. In contrast to
that, the linear loss time constant dominates the trap decay
because the initial steady state density is never too far inside
the quadratic loss regime. This effect is clearly observed in our
experiments. The above solutions are used to fit the decay and
loading data shown in fig. \ref{decay}. The relevant time
constants were 22.6~ms without repumping and 71.8~ms with
repumping. The parameter $\xi $ turned out to be negligible
without repumping ($\xi $ < 0.001) but takes the value $\xi $ =
0.32 when repumping is applied. The explicit expression for the
parameter $\xi $ in (\ref{xi}) shows that a diminishment of the
linear loss rate $\Gamma $ by a factor of 3.1 introduced by the
671~nm repumper yields the same increase in $\xi$ as an
enlargement of the capture rate $R$ by a factor of
$\left(3.1\right)^2\approx 10$. We can thus conclude that an order
of magnitude increase of the capture rate would yield a loss of
about 32 \% of the atoms due to inelastic collisions rather than a
transfer to the desired metastable state.

\section{Conclusion}

We have investigated the production of metastable calcium atoms in
a magneto-optical trap operating on the principal fluorescence
line. We obtain more than $10^{10}$ atoms in the 4$^3$P$_2$~state
per second at about 2 mK. An order of magnitude improvement
appears still possible by improving the MOT capture rate, however
at the price of entering the regime where inelastic light-induced
collisions begin to seriously hamper the production of
metastables.

\begin{acknowledgement}
This work has been partly supported by the Deutsche
Forschungsgemeinschaft under contract number DFG-He2334/2.3,
DAAD~415 probral/bu, and the European research network \em{Cold
Atoms and Ultra-precise Atomic Clocks} (CAUAC). We also
acknowledge the kind hospitality of the Schwerpunktprogramm SPP
1116.
\end{acknowledgement}


\begin{thebibliography}{99}
\bibitem{anderson} M.H. Anderson et al.: Science \textbf{269} (1995)

\bibitem{davis} K. Davis et al.: Phys. Rev. Lett. \textbf{75}, 3969-3973 (1995)

\bibitem{katori} T. Ido, Y. Isoya, and H. Katori: Phys. Rev. A, \textbf{61},
R061403 (2000)

\bibitem{PTB} T. Binnewies, G. Wilpers, U. Sterr, F. Riehle, J. Helmcke,
T.E. Mehlstäubler, E.M. Rasel, and W. Ertmer:
arXiv:physics/0105069 (May 2001)

\bibitem{NIST} E.A. Curtis, C.W. Oates, and L. Hollberg:
arXiv:physics/0104061 (April 2001)

\bibitem{spreeuw} R. Spreeuw, T. Pfau, U. Janicke, and M. Wilkens: Europhys. Lett
\textbf{32}, 469 (1995)

\bibitem{gruenert}  J. Grünert, G. Quehl, V. Elman, and A. Hemmerich: J. Mod. Opt. \textbf{47},
2733-2740 (2000)

\bibitem{derev} A. Derevianko: Phys. Rev. Lett. \textbf{87}, 023002 (2001)

\bibitem{PRL} J. Grünert, S. Ritter, and A. Hemmerich: to be published (2001)

\bibitem{kisters} T. Kisters, K. Zeiske, F. Riehle, and J. Helmcke: J. Appl. Phys. B
\textbf{59}, 89-98 (1994)

\bibitem{strumia} F. Strumia: In \textit{Laser Science and Technology}, edited by
A.N. Chester and S. Marellucci (Plenum 1988) pp. 367-401

\bibitem{oates} C. Oates, F. Bondu, R.W. Fox, and L. Hollberg: Eur. Phys. J. D \textbf{7},
449-460 (1999)

\bibitem{SrSr} K.R. Vogel, T.P. Dinneen, A. Gallagher, and J.L. Hall: IEEE
Trans. Instrum. Meas. , \textbf{48}, 618 (1999)

\end{thebibliography}
\end{document}